%% file: FanoSC_submit.tex
\documentclass[10pt,conference]{IEEEtran}

\usepackage{amssymb}
\usepackage{epsfig,verbatim}
\usepackage{algorithm}

\usepackage{algpseudocode}

\newcommand{\eqnlabel}[1]{\label{eqn:#1}}

\newcommand\blfootnote[1]{%
  \begingroup
  \renewcommand\thefootnote{}\footnote{#1}%
  \addtocounter{footnote}{-1}%
  \endgroup
}

\input{macros}

\usepackage{siunitx}

\usepackage{xcolor}

\usepackage{times}
\usepackage[cmex10]{amsmath}
\newcommand{\argmax}{\operatornamewithlimits{argmax}}
\newcommand{\argmin}{\operatornamewithlimits{argmin}}

\newtheorem{example}{Example}

\title{SC-Fano Decoding of Polar Codes}

\author{
\IEEEauthorblockN{
              Min-Oh Jeong  and Song-Nam Hong}
\IEEEauthorblockA{Ajou University, Suwon, Korea,\\
              email: \{jmo0802, snhong\}@ajou.ac.kr}
}

\begin{document}

\maketitle

\date{}

\blfootnote{}

%%%%%%%%%%%%%%%%%%%%%%%%%%%%%%%%%%%%%%%%%%%%%%%%%%%%%%%%%%%%%%%%%%%%%%%%%%%%%%%%%%%%%%%%%
\begin{abstract}
In this paper, we present a novel decoding algorithm of a polar code, named SC-Fano decoding, by appropriately incorporating the Fano sequential decoding into the standard successive-cancellation (SC) decoding. The proposed SC-Fano decoding follows the basic procedures of SC decoding with an additional operation to evaluate the reliability (or belief) of a current partial path. Specifically, at every decoding stage, it decides whether to move forward along a current path or move backward to find a more likelihood path. In this way, SC-Fano decoding can address the inherent drawback of SC decoding such as one wrong-decision will surely lead to a wrong codeword. Compared with the other improvements of SC decoding as SC-List (SCL) and SC-Stack (SCS) decodings, SC-Fano decoding has much lower memory requirement and thus is more suitable for hardware implementations. Also, SC-Fano decoding can be viewed as an efficient implementation of SC-Flip (SCF) decoding without the cost of cyclic-redundancy-code (CRC). Simulation results show that the proposed SC-Fano decoding significantly enhances the performance of SC decoding with a similar complexity as well as achieves the performance of SCL decoding with a lower complexity.
\end{abstract}

\begin{keywords}
Polar codes, successive-cancellation (SC) decoding, Sequential Decoding, SC-Fano Decoding.
\end{keywords}

%%%%%%%%%%%%%%%%%%%%%%%%

\section{Introduction}
 Polar codes, introduced by Arikan in \cite{Arikan2009}, achieve the symmetric capacity of the binary-input discrete memoryless channels (BI-DMCs) under successive-cancellation (SC) decoding. However, for practical finite lengths, polar codes with SC decoding yield poor performances compared with LDPC and Turbo codes.  In \cite{TalVardy2011},  SC list (SCL) decoding was developed, enabling to achieve the optimal maximum-likelihood (ML) performance with a sufficiently large list size. Despite its superior performance, SCL decoding suffers from high computational complexity, memory requirement, and lower decoding throughput.

 %%%%%%%%% Related Works (??) %%%%%%%%%
 
Recently in \cite{stack} and \cite{Hstack}, SC-Stack (SCS) decoding was proposed whose computational complexity can be very close to that of SC decoding in the high signal-to-noise ratio (SNR) regimes. In the operating SNR regimes (e.g., frame-error-rate (FER) $< 10^{-2}$), SCS decoding can have much lower decoding complexity than SCL decoding. Also, in \cite{Trifo2014} and \cite{Trifo2018}, an improved path-metric and efficient stack decoding algorithm for a polar code were developed. In contrast, SCS decoding requires larger space-complexity and its performance is not poor when the stack-size is small \cite{stack, Hstack}. Another improvement of SC decoding, called SCFlip (SCF) decoding, was proposed in  \cite{scflip}. Here, SC decoding is first performed to generate an initial estimated codeword. Then, it passes the cyclic-redundancy-check (CRC), the overall decoding is completed. Otherwise, SC decoding additionally proceeds by flipping a single information-bit which is carefully chosen with log-likelihood ratios (LLRs). SCF decoding terminates either when a valid codeword is found (i.e., CRC passes) or $T_{\rm max}$ additional trials are failed. Also, some improvements of SCF decoding were proposed in 
 \cite{imFlip}-\cite{Furkan2018}.

%%%%%% Our Contributions %%%%%

In this paper, we propose an alternative improvement of SC decoding, called SC-Fano decoding, by properly incorporating the Fano sequential decoding \cite{fano} into the standard SC decoding. Following the basic procedures of SC decoding, the proposed SC-Fano decoding has an additional step to evaluate the reliability (or belief) of a current path. Specifically, at every decoding stage, it decides whether to move forward along a current path or move backward to find a more likelihood path, which can overcome the major drawback of SC decoding such as one wrong-decision will destroy the overall decoding. As in the sequential decoding \cite{fano}, which is one of the low-complexity decodings for convolutional codes, the decision at every stage is made by comparing the (partial) path-metric (e.g., the path-metric of a current  path) with a dynamic threshold. As in the sequential decoding \cite{fano}, defining a path-metric plays a crucial role in the performance of SC-Fano decoding. The Fano metric, used in convolutional codes, is not appropriate for SC-Fano decoding due to the different code structures of polar and convolutional codes. We thus develop a novel (partial) path-metric suitable for the proposed SC-Fano decoding. Compared with the other improvements of SC decoding as SCL and SCS decodings, SC-Fano decoding has a lower memory requirement and thus is more suitable for hardware implementations. SC-Fano decoding can be viewed as an efficient implementation of the idea of SCF decoding without paying the cost of CRC. Namely, the former can flip an erroneous information bit immediately while the latter does it after decoding all the information bits. Finally, simulation results show that SC-Fano decoding enhances the performance of SC decoding with an almost same complexity and also achieves the performance of SCL decoding with a lower complexity.

%The sections of this letter are organized as follows. In Section \ref{sec:pre}, we briefly explain the polar codes and SC decoding. Section~\ref{sec:SCFano} describes the proposed SC-Fano decoding. In Section \ref{sec:simul}, we provide the simulation results to show the superiority of the proposed decoding. Conclusion is provided in section \ref{sec:con}.

%%%%%%%%%%%%%%%%%%%%%%%%%%

\section{Preliminaries}\label{sec:pre}

We briefly review the polar codes and the standard SC decoding in \cite{Arikan2009}. The polar codes are uniquely defined by the $(N,K,\Ac)$ where $N=2^n$ and $K$ represent the code length and the number of information bits, respectively, and  $\Ac \subseteq\{1,...,N\}$ denotes the information set which contains the indices of $K$ information bits. In \cite{Arikan2009}, the generator matrix of a polar code is obtained by $\mathbf\Gm_N=\Gm_2^{\otimes n}$ where $\Gm_2=\bigl[ \begin{smallmatrix} 1&0\\ 1&1 \end{smallmatrix} \bigr]$ and $\otimes$ represent the 2-by-2 Arikan kernel and the Kronecker product, respectively, and the $\Bm_N$ denotes the bit-reversal permutation matrix. Then, the polar encoding is performed as
\begin{equation} \label{encode}
 \mathbf{x}_1^N={\mathbf u_1^N}\Bm_N\Gm_N,
\end{equation}  where $\uv_{1}^{N}=(u_1,...,u_N)$ represents a message-vector containing both the information bits and zero frozen bits.

Next, we explain the standard SC decoding which produces the estimated information-bit $\hat{u}_i$ for $i=1,2,..,N$ in that order, from the 
observation $\yv_{1}^{N}=(y_1,y_2,...,y_N)$ such as
\begin{equation} \eqnlabel{u}
\hat{u}_i \equiv
\begin{cases}
0, & \mbox{if }i\in\Ac^c\\
\phi(\yv_{1}^{N}, \mathbf{\hat{u}}_1^{i-1}), & \mbox{if }i\in\Ac.
\end{cases}
\end{equation} Here, $\phi(\cdot)$ stands for the decision function  as 
\begin{equation} \eqnlabel{h}
\phi (\yv_{1}^{N},\mathbf{\hat{u}}_1^{i-1}) \equiv
\begin{cases}
0, & \mbox{if }\mathbf{\log}\left(\frac{\Pr(u_i=0|\yv_{1}^N,\mathbf{\hat{u}}_1^{i-1})}{\Pr(u_i=1|\yv_{1}^N,\mathbf{\hat{u}}_1^{i-1})}\right)\ge0\\
1, & \mbox{otherwise}
\end{cases} 
\end{equation}
where $\Pr(u_i|\yv_{1}^N,\mathbf{\hat{u}}_1^{i-1})$ (called branch-metric) is efficiently computed in a recursive way (see \cite{Arikan2009} for details).

%Note that the SC decoding sequentially estimates the $\hat{u}_i$'s in an ascending order.
%%%%%%%%%%%%%%%%%%%%%%%%%%%%

%%%%%%%%%%%%%%%%%%%%%%%%%%%%%%%%

\begin{figure}
\centering
\epsfig{file=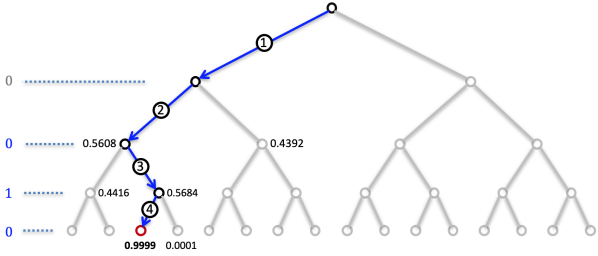, angle=0, width=\linewidth}
\caption{The example of SC decoding for the polar code with $N=4$ and $R=\frac{3}{4}$. Here, it estimates the wrong message-vector $\mathbf{\hat{u}}_1^4=(0,0,1,0)$. The number beside the node denotes the branch-metric $\Pr(\hat{u}_i|\mathbf{\hat{u}}_1^{i-1},\yv_1^N)$.}
\label{fig:sc}
\end{figure}

%%%%%%%%%%%%%%%%%%%%%%%%%%%%%%%%
\section{The Proposed SC-Fano Decoding}\label{sec:SCFano}

We propose a SC-Fano decoding which can improve the performance of SC decoding with an almost same complexity. The proposed decoding is developed by appropriately combining the Fano sequential decoding \cite{fano} with SC decoding. In detail, at every SC decoding stage, SC-Fano decoding decides whether to move forward along a current path or move backward to find a more likelihood path. Here, the decision is made by comparing the proposed partial path-metric of the current path with a dynamical threshold. Our major contribution is to present the (partial) path-metric suitable for SC-Fano decoding, which is different from the conventional Fano metric used in convolutional codes \cite{fano}. In the following subsections, we will explain the proposed path-metric and the detailed descriptions of SC-Fano decoding.

%%%%%%%%%%%%%%%%%%%%%%%%%%%%%%%%%%%%%%%%%%%
 \begin{figure}
\centering
\epsfig{file=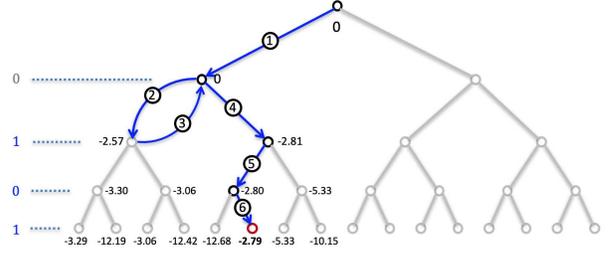, angle=0, width=\linewidth}
\caption{The example of SC-Fano decoding with $\Delta=3$ for the polar code with $N=4$ and $R=\frac{3}{4}$. Here,  it estimates the correct message-vector $\mathbf{\hat{u}}_1^4=(0,1,0,1)$. The number beside node denotes the path-metric $\Pc_{\rm SC-Fano}(\mathbf{\hat{u}}_1^{i})$.}
 \label{fig:fano}
\end{figure}

%%%%%%%%%%%%%%%%%%%%%%%%%%%%%%%%%%%%%%%%%%%%
\subsection{The proposed path-metric} \label{metric}

We propose a path-metric which will play a crucial role in SC-Fano decoding. As in the Fano sequential decoding \cite{fano}, the path-metric should satisfy the following two requirements:
%%%%%%%%%
\begin{enumerate}
\item It should be able to capture the reliabilities of the associated partial paths;
\item it should be properly normalized for the fair comparisons of the partial paths having different lengths.
\end{enumerate} 
%%%%%%%%%
First of all, there have been related studies focusing on the path-metric of SCS decoding (see \cite{Trifo2014} and \cite{Trifo2018} for details). Conventionally, the so-called Fano metric is employed for both the Fano sequential decoding and the stack decoding for conventional codes. In contrast, the path-metric in \cite{Trifo2014} and \cite{Trifo2018} is not suitable for the proposed SC-Fano decoding since it continuously decreases as the decoding-stage proceeds, which makes it to move backward frequently and thus can increase the decoding latency of SC-Fano decoding. Therefore, it is required to find a proper partial path-metric for SC-Fano decoding.

Toward this, since a path-metric used in SCL decoding \cite{TalVardy2011} satisfies the above requirement 1), we start with this metric defined as
\begin{equation}
\Pc_{\rm SCL}(\hat{\uv}_{1}^{i}) = \Pc_{\rm SCL}(\hat{\uv}_{1}^{i-1}) +  \log\left(\Pr(\hat{u}_i|\hat{\uv}_{1}^{i-1},\yv_{1}^{N})\right),
\end{equation} where note that this path-metric is computed recursively. Unfortunately, this metric cannot meet the requirement 2) because it tends to decrease steadily as a (decoded) path-length grows. In other words, as the decoding-stage proceeds, the resulting path-metric is more likely to decrease, equivalently, not to to pass the dynamic threshold $T$. Accordingly, the overall decoding process can be stuck (i.e., moving forward and backward repeatedly). We address this problem by introducing a proper normalization term which can account for the impact of the path-length on the reliability. Let $P_{e,j}$ denote an error-probability of the polarized (or synthesized) channel $i$. Note that it can be efficiently computed via density evolution \cite{Density}. Then, we first observe that for any length-$i$ path $\hat{\uv}_1^i=(\hat{u}_1,...,\hat{u}_i)$, the corresponding path-metric can be upper bounded as
\begin{equation}\label{eq:upper}
\Pc_{\rm SCL}(\hat{\uv}_{1}^{i}) \leq \sum_{j=1}^i \log{(1-P_{e,j})},
\end{equation} and equivalently, we have
\begin{equation}
\max_{\hat{\uv}_{1}^i \in \{0,1\}^i}\{\Pc_{\rm SCL}(\hat{\uv}_{1}^{i})\} \approx \sum_{j=1}^i \log{(1-P_{e,j})}.
\end{equation} Using this, we can satisfy the requirement 2 by normalizing the  path-metric such that the maximum  path-metric corresponding to a correct path would be $\log{1}= 0$ independently from the lengths of partial paths. Based on this, the proposed  path-metric is obtained as
\begin{equation} \label{metric}
\Pc_{\rm SC-Fano}(\hat{\uv}_{1}^{i}) = \Pc_{\rm SC-Fano}(\hat{\uv}_1^{i-1}) + \log\left(\frac{\Pr(\hat{u}_i|\hat{\uv}_{1}^{i-1},\yv_{1}^{N})}{1-P_{e,i}}\right),
\end{equation} with initial-metric $\Pc_{\rm SC-Fano}(\hat{\uv}_1^0)=\Pc_{\rm SC-Fano}(\hat{u_0})=0$. Note that the path-metric in  (\ref{metric}) is equal to
\begin{equation}
\Pc_{\rm SC-Fano}(\hat{\uv}_{1}^{i})=\log{\frac{\Pr(\mathbf{\hat u}_{1}^i |\yv_{1}^N)}{\prod_{j=1}^i (1-P_{e,j})}}.
\end{equation} Then, from (\ref{eq:upper}), we can confirm that 
\begin{equation}
\Pc_{\rm SC-Fano}(\hat{\uv}_{1}^{i})=\log{\frac{\Pr(\mathbf{\hat u}_{1}^i |\yv_{1}^N)}{\prod_{j=1}^i (1-P_{e,j})}} \leq 0
\end{equation} for any $i \in \{1,2,...,N\}$ since $\Pc_{\rm SCL}(\hat{\uv}_{1}^{i})=\log{\Pr(\mathbf{\hat u}_{1}^i |\yv_{1}^N)}$.

%%%%%%%%%%%%%%%%%%%%%%%%%%%%%%%%%%%%%%%%%%%
 \begin{figure}[t]
\centering
\epsfig{file=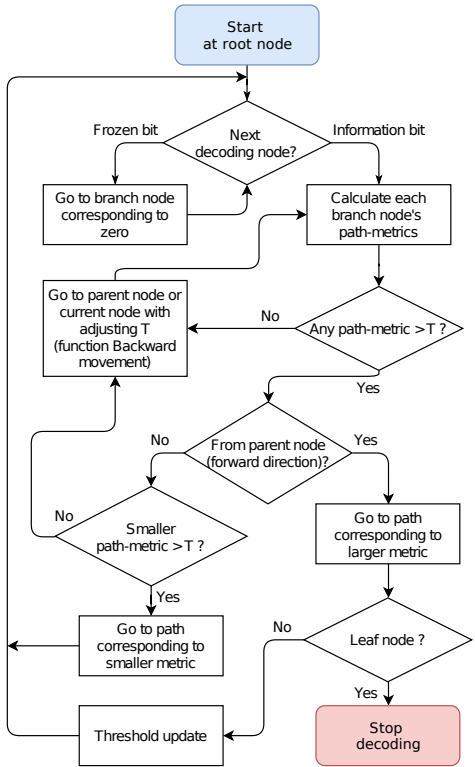, angle=0, width=180pt}
\caption{The flowchart of SC-Fano decoding.}
 \label{fig:flowchart}
\end{figure}

%%%%%%%%%%%%%%%%%%%%%%%%%%%%%%%%
\subsection{The SC-Fano decoding} \label{scfano}

As in SC, SCL, and SCS decodings, the proposed SC-Fano decoding is also performed based on the code tree as shown in Figs.~\ref{fig:sc} and~\ref{fig:fano}. We recall that as shown in Fig.~\ref{fig:sc}, SC decoding only moves forward (e.g., in the direction from the root node to leaf nodes) on the code tree such that at every decoding-stage, the path with a higher branch-metric is chosen among the two candidate paths. In contrast, SC-Fano decoding can move either forward or backward according to the corresponding path-metric and a current dynamic threshold $T$ (see Fig.~\ref{fig:fano}). The flowchart of the SC-Fano decoding is in Fig.~\ref{fig:flowchart}, and the detailed descriptions of the proposed SC-Fano decoding are provided in Algorithms 1, 2, and 3.

Algorithm 1 describes the main structure of SC-Fano decoding. Here, a binary variable $B$ indicates whether a current node is reached from a parent node (e.g., $B=0$) or a child node (e.g., $B=1$). For the case of $B=0$, the path with a higher path-metric (e.g., the good path) is chosen among the two candidate paths as in the standard SC decoding. Whereas, for the case of $B=1$, either the other path (e.g., the bad path) is selected or the process moves backward since in this case, the good-path has been already selected in a previous stage and the direction along that path was evaluated as not enough to be the most reliable path. 
Let $\gammav=(\gamma_1,...,\gamma_K)$ denote the length-$K$ binary vector where $\gamma_j = 1$ if the $j$-th information bit is decoded following the path with a lower branch-metric among two candidate paths, and $\gamma_j = 0$, otherwise. To avoid recomputing the  path metric of revisited paths, we introduce the length-$K$ vector  $\betav=(\beta_1,...,\beta_K)$ where $\beta_j$ contains the branch-metric for the decoding. 

Algorithm~\ref{move} describes the procedures of the backward movements in SC-Fano decoding where it is activated only when both path metrics in a forward direction are lower than a current threshold $T$. In Algorithm, the function $\Bc(\beta,j,T,\gamma)$ provides not only the index of an immediate parent information node but also the index of proper information node which is located upper level in the code tree then the immediate parent information node. This is because the frozen node is not used by SC-Fano decoding and thus it is possible for SC-Fano decoding to move backward more than one time. Furthermore, to make SC-Fano decoding more stable, the dynamic threshold $T$ is updated 
(see Algorithm~\ref{u}) whenever a path is firstly established. In algorithm 1, to automatically check whether a path is firstly established or note, it is compared with the path-metric of upper (parent) node (denoted by $\mu$). Without this update, the decoded path can be rotated permanently around near paths, which can yield an infinite decoding process. Also, via Algorithm~\ref{move} and Algorithm~\ref{u}, SC-Fano decoding can moderate the dynamic threshold $T$ automatically according to the reliability of the received signal.

 %The main frame of SC-Fano decoding algorithm is following the conventional Fano decoding algorithm \cite{fano}. Additionally, to provide readability on the SC-Fano decoding algorithm, the flowchart of the SC-Fano decoding is given in Fig.~\ref{fig:flowchart}.

We remark that in SC-Fano decoding, there is a parameter $\Delta$ to control the performance-complexity tradeoff. For example, the performance of SC-Fano decoding can be improved by setting a small value $\Delta$ while the corresponding computational complexity will increase as this tends to search more nodes in the code tree. Thus, when $\Delta$ is very small, SC-Fano decoding can achieve the performance of the optimal ML decoding. On the other hand,  if we set a larger value $\Delta$, the performance of SC-Fano decoding will be degraded while reducing the computational complexity. Namely, when $\Delta$ is very large, SC-Fano decoding is equivalent to the standard SC decoding. Likewise the advantage of the conventional Fano decoding over the stack decoding \cite{fano}, SC-Fano decoding has lower memory complexity than SCS decoding. Also, according to the previous works in \cite{stack} and \cite{Hstack}, SC-Fano decoding has also lower memory complexity than SCL decoding. In comparison with SC decoding,  SC-Fano decoding only requires the small amount of memories for partial paths' branch metrics $\betav$, indicator vector $\gammav$ and some values $B, T, \Delta$.

%%%%%%%%%%%%%%%%%%%%%%%%%%%%

\begin{algorithm}[t]
\caption{SC-Fano($\yv_1^N, \Delta, \Ac$)}\label{main}
\begin{algorithmic}[1]

%\Procedure{SC-Fano}{$\mathbf{Y}, \Delta, \alpha, \Ac$}
\\{\bf Initialization: } $i\gets1$, $j\gets0$, $B\gets0$
\State\hspace{0.1cm} $\betav=(\beta_1,...,\beta_K)=\zerov$, $\gammav=(\gamma_{1},...,\gamma_{K})= \zerov$

\While{$i\neq N+1$}
	\If{$i\in \Ac$}
  		 $m_{i,b} \eqdef \Pc(\mathbf{\hat{u}}_{1}^{i-1},\hat{u}_i=b)$ for $b\in \{0,1\}$ 
	        %\State $(M_i(\mathbf{\hat{u}}_{1}^{i-1},\hat{u}_i=0), M_i(\mathbf{\hat{u}}_{1}^{i-1},\hat{u}_i=1))\newline 
		%\hspace*{11.5em} \gets~${\bf Calcul\_M}($\mathbf{Y}, \mathbf{\hat{u}}_1^{i-1}, i, \alpha$)
		\If{$\max\{m_{i,0},m_{i,1}\}>T$}
			\If{$B=0$}
				 ${\hat{u}}_i=\argmax\{m_{i,0},m_{i,1}\}$
				\State $\beta_{j+1}= \max\{m_{i,0},m_{i,1}\}$, $\gamma_{j+1}=0$
				%\State $\gamma_{j+1}=0$
				\If{$j=0$} $\mu = 0$
					%\State $\mu\gets0$
				\ElsIf{$j\ne0$} $\mu= \beta_j$
					%\State $\mu\gets \mathbf{H}[j]$
				\EndIf
				\If{$\mu<T+\Delta$}  $T= \Uc(T, \Delta, \beta_{j+1})$
				\EndIf
				\State $i= i+1$, $j= j+1$
				%\State $j\gets j+1$
			\ElsIf{$B=1$}
				\If{$\min\{m_{i,0},m_{i,1}\}>T$}
					\State $\hat{u}_i=\argmin\{m_{i,0},m_{i,1}\}$
					\State $\beta_{j+1}= \min\{m_{i,0},m_{i,1}\}$, $\gamma_{j+1}=1$
					%\State $\gamma_{j+1}=1$
					\State $i= i+1$,  $j= j+1$, $B= 0$
					%\State $j\gets j+1$
					%\State $B\gets 0$
				\ElsIf{$\min\{m_{i,0},m_{i,1}\}\le T$}
					\If{$j=0$}
						 $T= T-\Delta$, $B= 0$
						%\State $B\gets 0$
					\ElsIf{$j\ne0$}
						\State $(T, j, B)= {\bf \Bc}(\betav, j, T, \gammav)$
						\State $i= \Ac(j+1)$
					\EndIf \EndIf \EndIf
		\ElsIf{$\max\{m_{i,0},m_{i,1}\}\le T$}
			\If{$j=0$} $T= T-\Delta$
				%\State $T\gets T-\Delta$
			\ElsIf{$j\ne0$}
				\State $(T, j, B)= {\bf \Bc}(\betav, j, T, \gammav)$, $i= \Ac(j+1)$
				%\State $i= \Ac(j+1)$
			\EndIf
		\EndIf
	\ElsIf{$i\notin \Ac$} ${\hat{u}}_i=0$, $i = i + 1$
  		%\State ${\hat{u}}_i\gets0$ and $i \gets i + 1$
  		%\State $i \gets i + 1$
	\EndIf
   \EndWhile\label{euclidendwhile}
   \State {Return} $\mathbf{\hat{u}}_1^N$
%\EndProcedure
\end{algorithmic}
\end{algorithm}
 
 % %%%%%%%%%%%%%%%%%%%%%%%%%%%%

\begin{algorithm}
\caption{Backward movement $\Bc$($\betav, j, T, \gammav$)}\label{move}
\begin{algorithmic}[1]
%\Procedure{$\Bc$}{$\mathbf{H}, j, T, \mathbf{I_b}$}
\While{True}
	\If{$j=1$} $\mu=0$
		%\State $\mu=0$
	\ElsIf{$j\ge2$} $\mu=\beta_{j-1}$
		%\State $\mu=\mathbf{H}[j-1]$
	\EndIf
	\If{$\mu\ge T$} $j = j-1$
		%\State $j\gets j-1$
		\If{$\gamma_{j+1}=0$}  $B =1$, Return $(T,j,B)$
		%$B\gets1$
			
			%\State {return} $(T,j,B)$
		\EndIf
	\ElsIf{$\mu<T$} $T= T-\Delta$, $B=0$
		%\State $B=0$
		\State {Return} $(T,j,B)$
	\EndIf
\EndWhile
%\EndProcedure
\end{algorithmic}
\end{algorithm}

%%%%%%%%%%%%%%%%%%%%%%%%%%%%

\begin{algorithm}
\caption{Threshold update $\Uc(T, \Delta, \tau)$}\label{u}
\begin{algorithmic}[1]
   \While{$T+\Delta < \tau$} $T= T+\Delta$
   %   \State $T\gets T+\Delta$
   \EndWhile
   \State {Return} $T$
\end{algorithmic}
\end{algorithm}

%%%%%%%%%%%%%%%%%%%%%%%%%%%%

%%%%%%%%%%%%%%%%%%%%%%%%%%%%%%%%

%%%%%%%%
\begin{example} In this example, we will show how the proposed SC-Fano decoding can outperform the standard SC decoding. For the comparison, we consider the polar code with the parameters  $(N=4, K=3, \Ac=\{2,3,4\})$. Assuming the message-vector $\uv_{1}^4=(0,1,0,1)$, the polar encoding in (\ref{encode}) generates the codeword $\xv_{1}^{4}=(0,1,0,1)$. Considering the additive-white-Gaussian-noise (AWGN) channel and BPSK modulation, the (noisy) received signal is given as  $\yv_1^4 = (1.4137, -1.5069, 2.3165, 1.3098)$. Figs.~\ref{fig:sc} and~\ref{fig:fano} illustrate the decoding procedures of  SC and SC-Fano decodings, respectively. We used the parameter $\Delta=3$ for SC-Fano decoding. This example shows that SC decoding fails to find a correct information-bit where $\mathbf{\hat{u}}_1^4=(0,0,1,0)$ is estimated. We notice that the failure of SC decoding is due to the wrong decision of $\hat{u}_2$ (e.g.,  \textcircled{\raisebox{-0.9pt}{2}} in  Fig. \ref{fig:sc}). In contrast, SC-Fano decoding can recognize this wrong decision as the path-metric is lower than the current threshold $T=-3$, namely, $\max\{\Pc(\hat{\uv}_{1}^2,\hat{u}_3 = 0), \Pc(\hat{\uv}_{1}^2,\hat{u}_3 = 1)\} = \max\{-3.30, -3.06\} = -3.06 < T=-3$. Here, note that the dynamic threshold was updated from $T=0$ to $T=-3$ because both path metrics $\Pc(\hat{\uv}_{1}^2=(0,0))=-2.57$ and $\Pc(\hat{\uv}_{1}^2=(0,1))=-2.81$ in Fig.~\ref{fig:fano} are lower than $T=0$. Algorithm \ref{move} makes to move backward  (e.g., \textcircled{\raisebox{-0.9pt}{3}} in Fig.~\ref{fig:fano}). Also, from Algorithm 1, the decoding process moves along the other path (\textcircled{\raisebox{-0.9pt}{4}} in the Fig. \ref{fig:fano}), which can eventually yield a correct information-bit $\hat{\uv}_{1}^{4}=(0,1,0,1)$.
\end{example}

%%%%%%%%%%%%%%%%%%%%%%%%%%%%%%%%%%%%%%%%%%%%%%%%%%
 \begin{figure}[t]
\centering
\epsfig{file=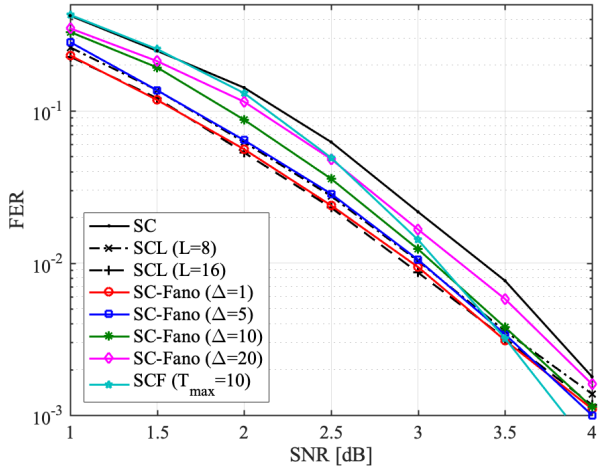, angle=0, width=\linewidth}
\caption{Performance comparisons of SC, SCL, SCF, and SC-Fano decodings for the length-128 polar code of the rate 
$R=\frac{1}{2}$.}
\label{fig:FER}
\end{figure}

%%%%%%%%%%%%%%%%%%%%%%
%%%%%%%%%%%%%%%%%%%%%%
\section{Simulation Results}\label{sec:simul}

\begin{figure}[t]
\centering
\epsfig{file=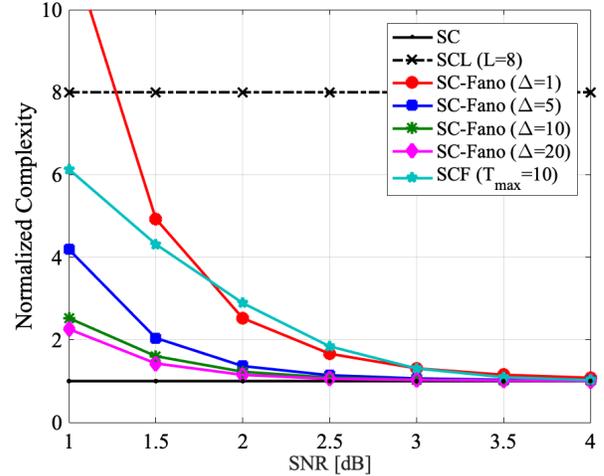, angle=0, width=\linewidth}
\caption{Complexity comparisons of SC, SCL, SCF, and SC-Fano decodings for the length-$128$ polar code of the rate $R=\frac{1}{2}$.}
\label{fig:complexity}
\end{figure}

We compare the proposed SC-Fano decoding with the SC, SCL and SCF decodings with respect to the frame-error-rate (FER) performance and the average decoding complexity. Regarding the complexity comparison, we used the notion of {\em normalized} complexity which implies that the complexities of SC, SCL, and SC-Fano decodings are normalized by that of SC decoding.  To be specific, the average normalized complexities of SC-Fano, SCL, and SCF decodings are respectively defined as $\chi_{\rm SC-Fano}=\EE[N_{SC-Fano}/N_{SC}]$, $\chi_{\rm SCL}=\EE[N_{SCL}/N_{SC}]$, and $\chi_{\rm SCF}=\EE[N_{SCF}/N_{SC}]$,
%\begin{equation}\label{eq:C-Fano}
%\chi_{\rm SC-Fano}=\EE\left[\frac{N_{SC-Fano}}{N_{SC}}\right]
%\end{equation} 
%\begin{equation}\label{eq:C-SCL}
%\chi_{\rm SCL}=\EE\left[\frac{N_{SCL}}{N_{SC}}\right]
%\end{equation} 
%\begin{equation}\label{eq:C-SCF}
%\chi_{\rm SCF}=\EE\left[\frac{N_{SCF}}{N_{SC}}\right],
%\end{equation} 
where $N_{SC}$, $N_{\rm SC-Fano}$, $N_{\rm SCL}$, and $N_{\rm SCF}$ denotes the summation of all the number of decoded bits during SC, SC-Fano, SCL, and SCF decodings, respectively. Note that $N_{\rm SC}$ is equal to the number of decoded bits (e.g., code length $N$).
For the simulations, AWGN channel and the polar code of length 128 and code rate $R=1/2$ are assumed.  For SC-Fano decoding, we considered the various parameters $\Delta=1,5,10,20$. Also, for SCL decoding, we used the list-sizes  $L=8, 16$. In fact, SCL decoding with $L=16$ almost achieves the optimal ML performance for this short-length polar code. The corresponding FER performances and decoding complexities are provided in  Figs.~\ref{fig:FER} and~\ref{fig:complexity}, respectively.

%%%%%%%%%%%%%%%%%%%%%%

From Figs.~\ref{fig:FER} and~\ref{fig:complexity}, we first confirm that the parameter $\Delta$ of SC-Fano decoding yields the performance-complexity tradeoff. Also, we observe that SC-Fano decoding with $\Delta=1$ can achieve the performance of SCL decoding with $L=16$ having a lower average computational complexity. Furthermore, the complexity reduction becomes a larger as $\SNR$ grows. At $\mbox{FER}=10^{-2}$, SC-Fano decoding with $\Delta=1$ can considerably reduce the complexity of SCL decoding by achieving the same performance and can significantly enhance the performance of SC decoding with a comparable complexity. In addition, we can see that SC-Fano decoding shows better performance than SCF decoding without the cost of additional CRC code where the 8-bit CRC and $T_{\rm max}=10$ are used for SCF decoding. Similarly shown in \cite{TalVardy2011}, the slope of SCF decoding is better than SC-Fano decoding as the concatenation of CRC code can improve the minimum distance of a polar code. We also confirmed that SC-Fano decoding together with bit-flipping idea in \cite{scflip} can outperform SCF decoding at all ranges of FERs, by enhancing the slope of SC-Fano decoding. The corresponding results are not included in this paper due to the lack of space.

%Therefore, we need to properly choose the increment $\Delta$ by taking the performance-complexity tradeoff into account.

%%%%%%%%%%%%%%%%%%%%%%%%%%%%%%%%%%%%%%%%%%%%%%%%%%
\section{Conclusion}\label{sec:con}

We proposed a novel decoding method of a polar code, named SC-Fano decoding, by incorporating the Fano sequential decoding into the standard SC decoding. The proposed SC-Fano decoding has much lower memory requirements than the other improvements as SCL and SCS decodings, and thus it is more suitable for hardware implementations. Via simulation results, it was shown that SC-Fano decoding improves the performance of SC decoding with a comparable complexity and also achieves the performance of SCL decoding with a lower decoding complexity. As an ongoing work, we will extend SC-Fano decoding for CRC-concatenated polar codes.

% Additionally, to improve the FER performance, it would be an interesting future work to develop the SC-Fano decoding for CRC concatenated polar code. 

%%%%%%%%%%%%%%%%%%%%%%%%%%%%%%

%%%%%%%%%%%%%%%%%%%%%%%%%%%%%%%%%%%%%%%%%%%

%%%%%%%%%%%%%%%%%%%%%%%%%%%%%%%%%%%%%%%%%%%%
\end{document}

%% file: macros.tex
\long\def\comment#1{}

\newfont{\bbb}{msbm10 scaled 700}

\newfont{\bb}{msbm10 scaled 1100}

\newcommand{\EE}{\mbox{\bb E}}

\newcommand{\uv}{{\bf u}}

\newcommand{\xv}{{\bf x}}
\newcommand{\yv}{{\bf y}}

\newcommand{\zerov}{{\bf 0}}

\newcommand{\Bm}{{\bf B}}

\newcommand{\Gm}{{\bf G}}

\newcommand{\Ac}{{\cal A}}
\newcommand{\Bc}{{\cal B}}

\newcommand{\Pc}{{\cal P}}

\newcommand{\Uc}{{\cal U}}

\newcommand{\betav}{\hbox{\boldmath$\beta$}}
\newcommand{\gammav}{\hbox{\boldmath$\gamma$}}

\newcommand{\SNR}{{\sf SNR}}

\newcommand{\eqdef}{\stackrel{\Delta}{=}}

%% file: FanoSC_submit.bbl
\begin{thebibliography}{1}


\bibitem{Arikan2009} E. Ar\i kan, ``Channel polarization: a method for constructing capacity-achieving codes for symmetric binary-input memoryless channels," {\em IEEE Trans.  Inf. Th.,} vol. 55, no. 7, pp. 3051-3073, Jul. 2009.

\bibitem{TalVardy2011} I. Tal and A. Vardy, ``List decoding of polar codes," {\em IEEE Trans. Inf. Th.,} vol. 61, no. 5, pp. 2213-2226, May 2015.

\bibitem{stack} K. Niu and K. Chen, ``Stack decoding of polar codes," {\em Elect. Letters,} vol. 48, no. 12, pp. 695-697, Nov. 2012.

\bibitem{Hstack} K. Chen, K. Niu and J. Lin, ``Improved successive cancellation decoding of polar codes," {\em IEEE Trans.  Commun.,} vol. 61, no. 8, pp. 3100-3107, Aug. 2013.

\bibitem{Trifo2014} V. Miloslavskaya and P. Trifonov, ``Sequential Decoding of Polar Codes," {\em IEEE Commun. Lett.,} vol. 18, no. 7, pp. 1127-1130, July 2014.

\bibitem{Trifo2018} P. Trifonov, ``A Score Function for Sequential Decoding of Polar Codes," in {\em Proc. IEEE Int. Symp. on Inf. Th. (ISIT),} pp. 1470-1474, Vail, CO, 2018.

\bibitem{scflip} O. Afisiadis, A. Balatsoukas-Stimming and A. Burg, ``A low-complexity improved successive cancellation decoder for polar codes," in {\em Proc. 48th Asilomar Conf.  Sig. Sys.  Comput.,}  pp. 2116-2120, Pacific Grove, CA, Nov. 2014.

\bibitem{imFlip} L. Chandesris, V. Savin and D. Declercq, ``An improved SCFlip decoder for polar codes," in {\em Proc. IEEE Global Commun. Conf. (GLOBECOM),}  pp. 1-6, Washington, DC, Dec. 2016.

\bibitem{Dflip} L. Chandesris, V. Savin and D. Declercq, ``Dynamic-SCFlip decoding of polar codes," {\em IEEE Trans.  Commun.,} vol. 66, no. 6, pp. 2333-2345, Jun. 2018.

\bibitem{flipDist} C. Condo, F. Ercan and W. J. Gross, ``Improved successive cancellation flip decoding of polar codes based on error distribution," in {\em Proc. IEEE Wire. Commun.  Net. Conf. Works. (WCNCW),}  pp. 19-24, Barcelona, Spain, Apr. 2018.

\bibitem{partitionedFlip} F. Ercan, C. Condo, S. A. Hashemi and W. J. Gross, ``Partitioned successive-cancellation flip decoding of polar codes," in {\em Proc. IEEE Int. Conf.  Commun. (ICC),} pp. 1-6, May 2018.

\bibitem{Furkan2018} F. Ercan, C. Condo and W. J. Gross, ``Improved bit-flipping algorithm for successive cancellation decoding of polar codes," {\em IEEE Trans.  Commun.}

\bibitem{fano} R. M. Fano, ``A heuristic discussion of probabilistic decoding," {\em IEEE Trans.  Inf. Th.,} vol. 9, no. 2, pp. 64-74, Apr. 1963.

\bibitem{Density} R. Mori and T. Tanaka, ``Performance of polar codes with the construction using density evolution," {\em IEEE Commun. Lett.,} vol. 13, no. 7, pp. 519-521, Jul. 2009.


\end{thebibliography}
